\documentclass[aps,prper,showpacs,twocolumn,floats,superscriptaddress,10pt]{revtex4-1}
\usepackage{graphicx}
\usepackage{dcolumn}
\usepackage{bm}
\usepackage{amssymb}
\usepackage{nicefrac}

\usepackage[latin3]{inputenc}
\usepackage[makeroom]{cancel}
\usepackage{amsmath}
\usepackage{amsfonts}
\usepackage{amssymb}
\usepackage{chngcntr}
\usepackage{color}
\usepackage[left=2cm,right=2cm,top=2cm,bottom=2cm]{geometry}

\usepackage{graphicx} 
\usepackage{dcolumn}
\usepackage{bm}
\usepackage{simplewick}
\usepackage{array}
\usepackage{appendix}

\RequirePackage[
   hyperindex,colorlinks,bookmarksnumbered,
   plainpages=true,pdfstartview=FitH]{hyperref}
\hypersetup{linkcolor=blue,urlcolor=blue,citecolor=blue}
\usepackage{hyperref}

\definecolor{purple}{rgb}{0.5,0,0.6}

\usepackage{ulem}

\begin{document}

\title{Thermoelectric Transport in a Three-Channel Charge Kondo Circuit}
\date{\today}
\author{T. K. T. Nguyen}
\affiliation{Institute of Physics, Vietnam Academy of Science and Technology,
10 Dao Tan, Hanoi, Vietnam}
\author{M. N. Kiselev}
\affiliation{The Abdus Salam International Centre for Theoretical Physics, Strada
Costiera 11, I-34151, Trieste, Italy}

\begin{abstract}{\color{black}
We investigate theoretically the thermoelectric transport through
a circuit implementation of the three-channel {\color{black} charge} Kondo model {\color{black} quantum simulator}
[Z. Iftikhar \textit{et al.}, Science \textbf{360}, 1315 (2018)].
The universal temperature scaling law of the Seebeck coefficient is computed perturbatively {\color{black} approaching} the non-Fermi liquid strong coupling fixed point using {\color{black} abelian} bosonization technique. The predicted {\color{black} $T^{1/3}\log T$} scaling behaviour of the thermoelectric power sheds a light on the properties of $Z_3$ emerging parafermions and gives an access to exploring {\color{black} pre-}fractionalized zero modes in the quantum transport experiments. We discuss a generalization of approach for investigating a multi-channel Kondo problem with emergent $Z_N\to Z_M$  {\color{black} crossovers} between {\color{black} ``weak"} non-Fermi liquid regimes corresponding to different low-temperature fixed points.}   
\end{abstract}


\maketitle

{\color{black} Quantum thermoelectricity is one of the most rapidly developing directions of the quantum technology \cite{Whitney, Zlatic}. Modern progress in fabrication of nano-devices operating at ultra low (milli-Kelvin range) temperatures opens an access to a broad variety of the charge, spin and heat transport phenomena governed entirely by the quantum effects \cite{Blanter,kisbook}.
In particular, quantization effects in behaviour of quantum simulators (see. e.g. \cite{quant_1,quant_2,quant_3,quant_4,quant_5,quant_6,2CK_experiment_nature2015,
3CK_experiment_science2018}) at the regimes affected by  quantum criticality are challenging for both experimental and theoretical communities. 

Among a large variety of available quantum devices, Quantum Dots (QD) {\color{black} play an important} and significant role. On the one hand, the QD devices \cite{Blanter,kisbook} are highly controllable and fine-tunable setups operating at the regimes adjustable by external electric and magnetic fields at both weak and strong out of equilibrium conditions. On the other hand, the QD devices as the quantum impurity simulators provide an important playground for understanding the influence of strong electron-electron interactions, interference effects and resonance scattering on the quantum transport.

One of the cornerstone effects
showing both the resonance scattering and strong interactions as two sides of the same coin is the Kondo effect \cite{Kondo_1964,Hewson}. While conventional Kondo phenomenon is attributed to a spin degree of freedom of the quantum impurity \cite{TW1983,AFL1983,glaz_ko}, the unconventional charge Kondo effect is dealing with an iso-spin implementation
of the charge quantization \cite{Flensberg,Matveev1995,Furusaki_Matveev,aleiner_glazman,LeHur,LeHur_Seelig}. Kondo model \cite{Hewson,TW1983,AFL1983} is one of known realizations of the
``minimal models" archetypal for description of both Fermi liquid (FL) and non-Fermi liquid (NFL) regimes associated with the collective many-body phenomena. 

The FL paradigm is one of the most important achievements of  twentieth-century condensed matter physics \cite{Landau}. It provides a tool to account for the effects of interaction in the equilibrium and out-of-equilibrium correlation functions \cite{Coleman}. While FLs are well defined objects characterized by some universal properties of corresponding quantum field theory encoded in scaling behaviour of the correlation functions \cite{Coleman} or, equivalently, certain constrains in the phenomenological description, NFLs represent rather {\it ``terra incognita"} unless some strong fingerprints of the quantum behaviour inconsistent with the FL paradigm directly follow from known classes of the models.
Fortunately, the multi-channel Kondo (MCK) model gives an access to collective behaviour 
completely different from the FL theory predictions \cite{Nozieres_Blandin_1980, Cox1998, Affleck_Ludwig_1993}. The beauty 
and ``simplicity" of the Kondo model makes it attractive for both experimental implementation of the strongly correlated physics and theoretical benchmarking 
of the many-body approaches beyond conventional mean-field or perturbation theory techniques. The price one has to pay for using minimal model is in immense complications in experimental fabrication of the MCK devices \cite{Goldhaber-Gordon_Nature_2007} and necessity to use advanced and cumbersome theoretical tools for the description of the strong coupling regimes \cite{book_bosonization_1998,giamarchibook,seneshal}. 

Recently, the breakthrough experiments \cite{2CK_experiment_nature2015,3CK_experiment_science2018} convincingly demonstrated the paramount importance of the MCK physics for the quantum charge transport through the nano-device. The few-channel Kondo physics is shown to be extended beyond existing realization of a two-channel Kondo (2CK) effect \cite{oreg_gg,Goldhaber-Gordon_Nature_2007} to a three-channel Kondo (3CK) phenomenon. While NFL regime of 2CK \cite{AD1984,FGN_1,FGN_2} is explained by an emergent $Z_2$ symmetry attributed to Majorana fermions \cite{Emery_Kivelson, book_bosonization_1998}, the 3CK Kondo physics is known to be associated with $Z_3$ parafermion states \cite{z3_0,z3_1,z3_2,z3_3,zn_1,zn_2}. 

In this {\it Letter} we address a fundamental question on how the NFL physics of 3CK model influences the quantum thermoelectric transport through the 
quantum simulators reported in \cite{2CK_experiment_nature2015,3CK_experiment_science2018}. In particular, we investigate theoretically
a scaling behaviour of thermoelectric coefficients 
and  analyse {\color{black} crossovers} between the NFL regimes associated with different low temperature strong coupling fixed points of 3CK. {\color{black}  The temperature scaling of thermopower is closely related to corresponding scaling of the fundamental quantum thermodynamic quantities (see \cite{ent2}) providing (as opposed to electric conductance measurements \cite{2CK_experiment_nature2015,3CK_experiment_science2018}) an access to fractionally quantized entropy \cite{ent2}.}}

\begin{figure}[h]
\includegraphics[width=65mm,angle=0]{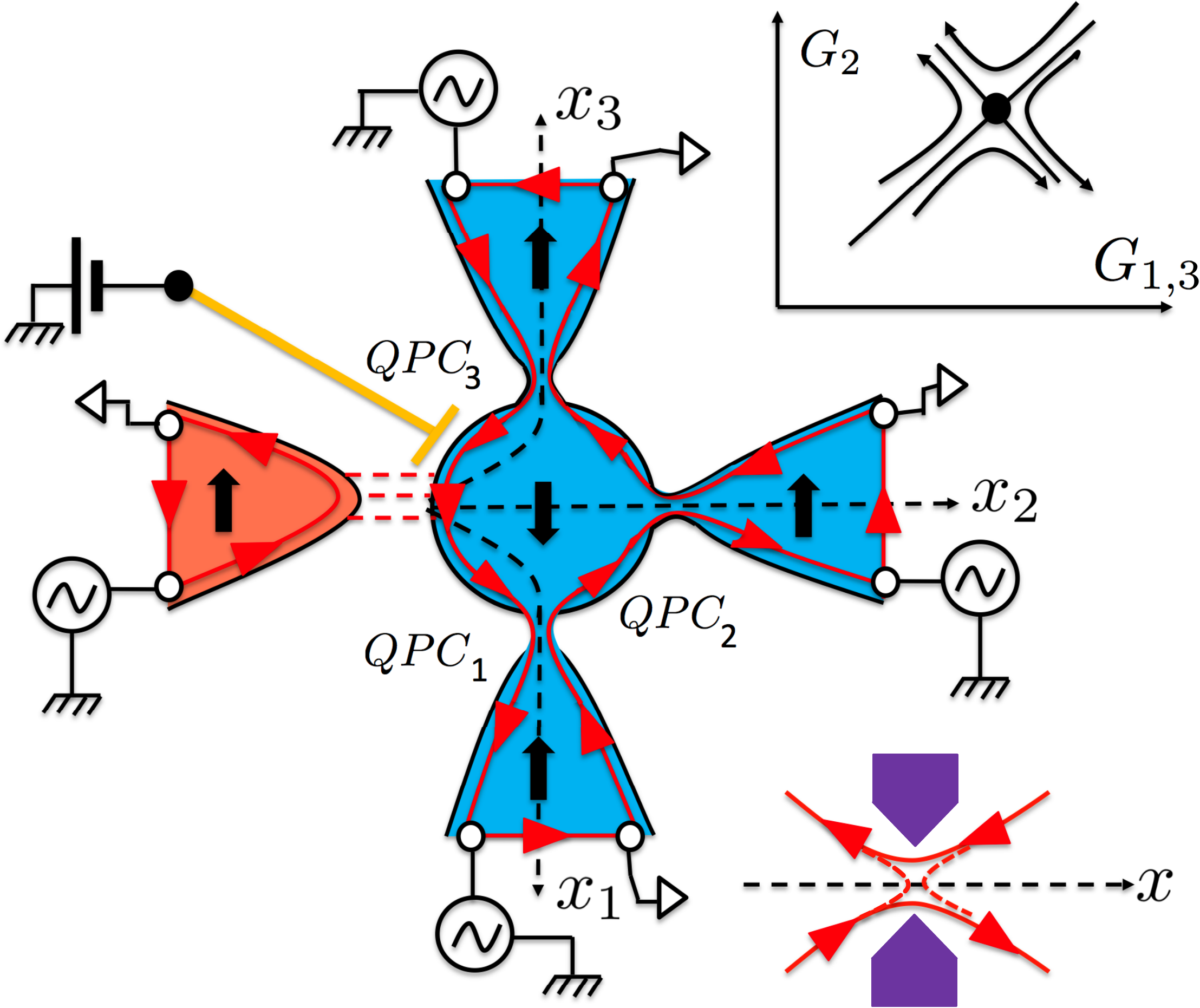}
\caption{(Color online) Schematic of three channel charge Kondo (3CK) setup. Central metallic island, aka Quantum Dot  (QD) is connected to four electrodes formed by two dimensional electron gas (2DEG). The state in QD is characterized by
the iso-spin $\sigma$$=$$\downarrow$. The states in the electrodes are characterized by the iso-spin $\sigma$$=$$\uparrow$. The left (orange) electrode is heated to the temperature $T$$+$$ \Delta T$ and connected to the rest of the setup through a tunnel contact (red dashed lines). The reference temperature of the QD and three blue electrodes is $T$. The yellow plunger gate is used to control a mesoscopic Coulomb blockade in the QD. The setup is fine tuned by external magnetic field to the Integer Quantum Hall (IQH) regime $\nu$$=$$2$. The current propagates along spin-polarized edge channels (red solid lines with arrows). Only one relevant (outer) chiral edge channel is shown. 
The transparencies 
of the quantum point contacts
QPC$_1$-QPC$_3$ {\color{black} (narrow blue constrictions)} are controlled by the surface split gates {\color{black} (magenta boxes in insert)}. {\color{black} Black dashed lines depict three independent $x$-axes with origins located in the middle of constrictions (QD boundary). Zoomed in edge state at one of the QPCs 
and weak backscattering (red dotted lines)
are shown in the lower insert. 
} The identical thermo-voltages are applied across the tunnel contacts to nullify the net electric current through the device. {\color{black} Upper} insert shows schematically a renormalization group flow for {\color{black} 3CK}. The unstable strong coupling fixed point at $G_1$$=$$G_2$$=$$G_3{\color{black} \approx} 0.69e^2/h$ corresponds to the 3CK non-Fermi-liquid regime.}
\label{f1} 
\end{figure}

\textit{Model} -- In a nano-device (see Fig. \ref{f1}) designed
to be used for thermoelectric measurements \cite{thermo_exp1,thermo_exp2,thermo_exp3,thermo_exp4},
the drain consists of a large metallic QD electrically
connected to two-dimensional electron gas (2DEG) electrodes through  
three quantum point contacts (QPCs) as proposed in Refs. \cite{2CK_experiment_nature2015,3CK_experiment_science2018}.
The 2DEG is in the integer quantum Hall (IQH) regime at the filling
factor $\nu=2$. The QPCs are fine-tuned to satisfy condition that
only the outer {\color{black} spin polarized} chiral edge current
is partially transmitted across the QPCs. The drain is at the reference
temperature $T$. The source is separated from the QD by a tunnel
barrier with low transparency $|t|\ll1$ as described by a tunnel
Hamiltonian $H_{tun}=\sum_{k}(tc_{k}^{\dagger}d+\text{h.c.})$ with
$c$ and $d$ denote the electrons in the left lead and in the dot.
{\color{black} The temperature of the source can be controlled by the ``floating island"
technique \cite{quant_1}. A micron-sized metal island \cite{quant_1} 
is electrically connected by several channels at 
opposite voltages (to have a zero dc voltage) in the left electrode upstream to the tunnel contact to the Kondo island \cite{fp_private}. 
Electrons in the ``floating island" are heated up with Joule heat. The resulting temperature is measured by the noise-based thermometry \cite{quant_1,alter_heat,en_rel,fp_private}.}
The temperature difference $\Delta T$ across the tunnel barrier is
assumed to be small compared to the reference temperature $T$ to
guarantee the linear response regime for the device at the weak link
\cite{nkk_2010}. The central metallic island (QD) is in a regime
of weak (mesoscopic) Coulomb blockade \cite{aleiner_glazman, gg_2011} {\color{black} characterized by the charging energy $E_C$}. The gate voltage $V_{g}$ is
used to tune charge degeneracy $N(V_{g})$ to the regimes of Coulomb
peaks ($N$ is half-integer) and Coulomb valleys ($N$ is integer).
The Kondo physics is observed through the measurements of the QPCs
differential conductances {\color{black} $G_\alpha$} at zero bias voltages $V_\alpha\to 0$ {\color{black} through the measurement of $I_\alpha/ V_\alpha$}
\cite{3CK_experiment_science2018} 
(see Fig. \ref{f1}). The MCK regime is fine-tuned by setting
transmission coefficients across QPCs to be equal. Applying a thermo-voltage 
{\color{black} $\Delta V_{th}$} to implement a zero-current condition for the electric current between
the source (orange lead) and drain (QD and three blue leads) allows to access the thermoelectric {\color{black} coefficient
$G_{T}$ through the measurements of $I_\alpha/\Delta T$ and Seebeck coefficient aka thermopower (TP) $S=G_{T}/G|_{I=0}=-\Delta V_{th}/\Delta T$ \cite{thermo_exp1,thermo_exp1}.} 

The mapping of IQH setup to a MCK problem is explained in details
in Ref. \cite{nk2018}. 
We assign the iso-spin
$\uparrow$ to the electrons in each QPC and the iso-spin $\downarrow$
to the electrons in the QD. 
{\color{black} The charge iso-spin flips when the
electrons move in- and out- of the QD. Backscattering transfers ``moving in-" the QD electrons to ``moving out-" from the QD electrons and vice versa.} The number of QPCs is equivalent to the number of orbital channels in the conventional $S$$=$$1/2$ Kondo problem.

It is convenient to describe the interacting electrons in the QD and QPCs in the bosonized representation \cite{Matveev1995,Furusaki_Matveev,aleiner_glazman,LeHur,LeHur_Seelig,MA_theory}.
We start with the Euclidean action $S=S_{0}+S_{C}+S^{\prime}$ describing QD
and three QPCs. The action $S_0$ \cite{seneshal} stands for the free part representing three copies of free one-dimensional electrons in QPC$_\alpha$:
\begin{eqnarray}
S_{0}=\frac{v_{F}}{2\pi}\!\sum_{\alpha=1}^{3}\int_{0}^{\beta}\!dt\int_{-\infty}^{\infty}\!dx\!\left[\frac{\left(\partial_{t}\phi_{\alpha}(x,t)\right)^{2}}{v_{F}^{2}}+\left(\partial_{x}\phi_{\alpha}(x,t)\right)^{2}\right],\;\;\;\;\nonumber
\label{act0}
\end{eqnarray}
Here $\phi_\alpha(x,t)$ denotes bosonic field describing the transport through 
QPC$_\alpha$ (see also \cite{nk2018}) and $v_{F}$ is a Fermi velocity \cite{comaxes}, 
{\color{black} $\beta$$=$$1/T$ (we adopt the units $\hbar$$=$$c$$=$$k_B$$=$$1$)}. 

{\color{black} The effects of the weak mesoscopic Coulomb blockade in the QD are described by the Hamiltonian $H_C=E_C[\hat n -N(V_g)]^2$.}
{\color{black} In the spirits of Andreev-Matveev theory \cite{MA_theory}, 
the operator $\hat{n}$  in the Hamiltonian $H_C$ accounts for the electrons entering the dot through the left weak tunnel barrier and three QPCs ($\hat{n}$$=$$\hat{n}_L $$+$$\hat{n}_{QPC}$). 
The number of electrons entering QD from the QPCs is related to the bosonic fields
$\phi_\alpha$ as $\hat{n}_{QPC}\to\sum_{\alpha=1}^{3}\phi_\alpha (0,t)/\pi$ \cite{MA_theory,comcb}, while the operator $\hat{n}_L$ counting the number of electrons tunneling from the left electrode can be replaced by the function $n_{\tau}(t)=\theta(t)\theta(\tau-t)$ \cite{MA_theory}. Here $\theta(t)$ is the unit step function (Heaviside function.)}
The Coulomb blockade action $S_{C}$ in bosonized representation \cite{Matveev1995,Furusaki_Matveev,aleiner_glazman,LeHur,LeHur_Seelig,MA_theory} is given by:
\begin{eqnarray}
S_{C}\left(\tau\right)=\int_{0}^{\beta}\!\!dtE_{C}\left[{\color{black} n_\tau(t)}+\frac{1}{\pi}\sum_{\alpha=1}^3\phi_{\alpha}(0,t)-N(V_{g})\right]^{2}.\;\;\;\;\nonumber
\end{eqnarray} 
Finally, the action $S^{\prime}$ 
\begin{eqnarray}
S^{\prime} & = & -\frac{D}{\pi}\;\sum_{\alpha=1}^{3} {\color{black}|r_\alpha|}\int_{0}^{\beta}dt \cos\left[2\phi_{\alpha}(0,t)\right]\;\;\;\nonumber
\end{eqnarray}
characterizes the backscattering at QPCs with $r_{\alpha}$ is
reflection amplitude for the QPC$\alpha$ and $D$ is the bandwidth (ultraviolet cutoff). We consider the symmetric situation, where 
$|r_{1}|$$=$$|r_{2}|$$=$$|r_{3}|$$\equiv|r|$$\ll $$1$.

\textit{Three {\color{black} normal modes}} -- we introduce three linear combinations
of the fields $\phi_{\alpha}$ to represent charge, {\color{black} pseudo-spin} and flavour modes (see, e.g. \cite{4CK_Fabrizio}): 
\begin{eqnarray}
\phi_{c}\left(x,t\right) & = & \frac{1}{\sqrt{3}}\left(\phi_{1}\left(x,t\right)+\phi_{2}\left(x,t\right)+\phi_{3}\left(x,t\right)\right),\nonumber \\
\phi_{s}\left(x,t\right) & = & \frac{1}{\sqrt{2}}\left(\phi_{1}\left(x,t\right)-\phi_{3}\left(x,t\right)\right),\nonumber \\
\phi_{f}\left(x,t\right) & = & \frac{1}{\sqrt{6}}\left(\phi_{1}\left(x,t\right)-2\phi_{2}\left(x,t\right)+\phi_{3}\left(x,t\right)\right),\label{eq:csf}
\end{eqnarray}
and the same for the dual boson fields $\frac{1}{\pi}\partial_{x}\theta_{\alpha}$$=$$\Pi_{\alpha}$$=$$-\frac{1}{v_{F}}\partial_{t}\phi_{\alpha}$
satisfying {\color{black} equal-time} commutation relations: $[\phi_{\alpha}(x),\Pi_{\alpha'}(x')]$$=$$i\,\delta(x-x')\delta_{\alpha\alpha'}$ \cite{book_bosonization_1998,giamarchibook,seneshal}.
Here $\alpha$,$\alpha^{\prime}$ denote charge, {\color{black} pseudo-spin $S$$=$$1$}, and flavour.
The {\color{black} pseudo-spin} and flavour modes are {\color{black} related} to two diagonal Gell-Mann matrices of SU($3$) group \cite{kisbook,com_2}. {\color{black} The parametrization (\ref{eq:csf}) explicitly breaks the symmetry between QPCs while
this symmetry is preserved in the model. We therefore need to use two additional parametrizations \cite{GMb} corresponding to the cyclic permutations of the indices $1$$\to $$2$$\to $$3$ (re-numeration of the QPCs) and apply a symmetrization procedure at the point $|r_\alpha|$$=$$|r|$.
For brevity we omit index labelling the representation \cite{GMb} in the notations.} 

The action in the charge, {\color{black} pseudo-spin} and flavour modes {\color{black} (for illustration we use (\ref{eq:csf}) $\vec\phi_{csf}$$\equiv $$\vec\phi^\mu$ \cite{GMb})} is 
\begin{eqnarray}
&&S_{0}=\!\frac{v_{F}}{2\pi}\!\!\int_{0}^{\beta}\!\!\!\!dt\!\int_{-\infty}^{\infty}\!\!\!\!\!\!dx\!\!{\sum_{\alpha=c,s,f}}\!\!\left[\!\frac{\left(\partial_{t}\phi_{\alpha}(x,t)\right)^{2}}{v_{F}^{2}}\!+\!\left(\partial_{x}\phi_{\alpha}(x,t)\right)^{2}\!\right]\!\!,\;\;\\\label{eq:S0csf}
&&S_{C}\left(\tau\right)=\int_{0}^{\beta}\!\!dtE_{C}\left[{\color{black} n_\tau(t)}+\frac{\sqrt{3}}{\pi}\phi_{c}(0,t)-N(V_{g})\right]^{2},\\\label{eq:SCcsf}
\! &&S^{\prime}  =  -\frac{D}{\pi}|r|\!\!\!\int_{0}^{\beta}\!\!\!dt\left\{ \cos\left[\frac{2}{\sqrt{3}}\phi_{c}(0,t)-\frac{2\sqrt{2}}{\sqrt{3}}\phi_{f}(0,t)\right]\right.\nonumber \\
 && \left.+2\cos\!\left[\frac{2}{\sqrt{3}}\phi_{c}(0,t)+\frac{\sqrt{2}}{\sqrt{3}}\phi_{f}(0,t)\right]\cos\!\left[\sqrt{2}\phi_{s}(0,t)\right]\!\right\}.\;\;\;\;\;\;\label{eq:SBScsf}
\end{eqnarray}

Action $S_{0}$ is particle-hole (PH) symmetric. PH transformation
in the action $S_{C}$ corresponds to change $N$ to $-N$ (electrons
are replaced by holes). As a result, the transport coefficients $G$
and $G_{T}$ transform under PH transformation as follows: $G(N)=G(-N)$
and $G_{T}(-N)=-G_{T}(N)$. Besides, the thermoelectric transport
requires breaking of the particle-hole symmetry described by the backscattering
action $S^{\prime}$.

Furthermore, due to Coulomb blockade effect all transport coefficients
are periodic in $N(V_{g})$ and the action is invariant with respect
to the shift $N\to N+1$. To show it we notice that the electron travels
from/to QD to/from one of the QPCs. In the setup (see Fig. \ref{f1}), there
are three possible ways to do it: i) electron enters QD from the QPC1:
$\phi_{1}\rightarrow\phi_{1}+\pi$, $\phi_{2}\rightarrow\phi_{2}$,
$\phi_{3}\rightarrow\phi_{3}$. As a result $\phi_{c}\rightarrow\phi_{c}+\nicefrac{\pi}{\sqrt{3}}$,
$\phi_{s}\rightarrow\phi_{s}+\nicefrac{\pi}{\sqrt{2}}$, $\phi_{f}\rightarrow\phi_{f}+\nicefrac{\pi}{\sqrt{6}}$;
ii) electron enters QD from the QPC2: $\phi_{1}\rightarrow\phi_{1}$,
$\phi_{2}\rightarrow\phi_{2}+\pi$, $\phi_{3}\rightarrow\phi_{3}$
then $\phi_{c}\rightarrow\phi_{c}+\nicefrac{\pi}{\sqrt{3}}$, $\phi_{s}\rightarrow\phi_{s}$,
$\phi_{f}\rightarrow\phi_{f}-\nicefrac{2\pi}{\sqrt{6}}$; and iii)
electron enters QD from the QPC3: $\phi_{1}\rightarrow\phi_{1}$,
$\phi_{2}\rightarrow\phi_{2}$, $\phi_{3}\rightarrow\phi_{3}+\pi$
then $\phi_{c}\rightarrow\phi_{c}+\nicefrac{\pi}{\sqrt{3}}$, $\phi_{s}\rightarrow\phi_{s}-\nicefrac{\pi}{\sqrt{2}}$,
$\phi_{f}\rightarrow\phi_{f}+\nicefrac{\pi}{\sqrt{6}}$. These discrete
transformations keep the backscattering action $S^{\prime}$ invariant
and increase charge of the QD by one. We rely upon these transformations
{\color{black} (as well as corresponding transformations in basis $\vec\phi^\lambda$ and $\vec\phi^\rho$ \cite{GMb})}
in the perturbative calculations (see details in 
\cite{suppl}).

\textit{Perturbative calculations} -- The transport coefficients
$G$ and $G_{T}$ are expressed in terms of the correlation function $K(\tau)$ \cite{MA_theory}:
\begin{eqnarray}
K(\tau) & = & Z(\tau)/Z(0),\nonumber \\
Z(\tau) & = & \int\exp[-S_{0}-S_{C}(\tau)-S^{\prime}]\prod_{\alpha}\mathcal{D}\phi_{\alpha}(x,t).\label{zz}
\end{eqnarray}
This correlation function is characterized by the following symmetries
associated with PH and shift transformation: $K$$($$\beta$$-$$\tau,$$N$$)$$=$$K$$($$\tau,$$1$$-$$N$$)$ and $K$$($$\beta$$-$$\tau,$$N$$)$$=$$K$$($$\tau$$,$$-$$N$$)$.

The electric conductance $G$ \cite{Furusaki_Matveev} is given by 
\begin{eqnarray}
G & = & \frac{G_{L}\pi T}{2}\int_{-\infty}^{\infty}\frac{1}{\cosh^{2}(\pi Tt)}K\left(\frac{1}{2T}+it\right)dt~.\label{elec_cond_def}
\end{eqnarray}
Here $G_{L}\ll e^{2}/h$ denotes the tunnel conductance of the left
barrier calculated ignoring influence of the dot. The thermoelectric coefficient $G_T$
takes the form \cite{MA_theory} 
\begin{eqnarray}
\!\!\!G_{T}=-\frac{i\pi^{2}}{2}\frac{G_{L}T}{e}\!\!\int_{-\infty}^{\infty}\!\!\frac{\sinh(\pi Tt)}{\cosh^{3}(\pi Tt)}K\left(\frac{1}{2T}+it\right)dt~.\label{thercond_def}
\end{eqnarray}

The correlator $K(\tau)$ acquires a simple form in the absence of
the backscattering. The action $S_{0}+S_{C}$ is Gaussian and the
functional integrals are explicitly evaluated resulting in \cite{footnote1}
(see details of calculations in \cite{suppl}).
\begin{eqnarray}
K(\tau)\left|_{r=0}\right.=K^{(0)}(\tau)=\left[\frac{\pi^{2}T}{3\gamma E_{C}}\frac{1}{|\sin\left(\pi T\tau\right)|}\right]^{\frac{2}{3}}.\label{ktau}
\end{eqnarray}

Here $\gamma$$=$$e^{\bf C}$$\approx $$1.78$, ${\bf C}$$\approx $$0.577$. The backscattering $r\neq0$ explicitly breaks the PH symmetry. However,
the mechanism of the PH symmetry breaking is different for the FL  ($M=1$) and MCK-NFL, ($M\geq2$) states.
Namely, for the FL case, there exists only one gapped mode associated
with the charge. Therefore, the PH symmetry breaking occurs already
in the first order of the perturbation theory \cite{MA_theory}. If, however,
there are $M-1$ gapless modes describing spin and flavours for the
MCK-NFL, the first order perturbative correction vanishes and PH symmetry
breaking occurs in the second order. The non-vanishing contribution
to the $G_{T}$ and $S$ is associated with the fluctuations of $M-1$
gapless modes. We process with the perturbative calculations at the
second order $K^{(2)}(\tau)=K_{C}(\tau)\left(\langle S^{\prime2}\rangle_{\tau}-\langle S^{\prime2}\rangle_{0}\right)/2$.
The validity of the perturbation theory at $|r|^{2}\ll1$ for 2CK \cite{MA_theory} 
is justified by the condition for the temperature regime $T^{\ast}\ll T\ll E_{C}$
where $T^{\ast}=|r|^{2}E_{C}$ \cite{MA_theory}. We refer to this regime
as the \textit{weak {\color{black} NFL} regime}.

\textit{Scaling of transport coefficients} -- The main contribution
to the electric conductance does not depend on $|r|$.
Its temperature scaling is fully determined by the form of $K^{(0)}(\tau)$
given by Eq.(\ref{ktau}) (see \cite{footnote2,footnote3}): 
\begin{eqnarray}
G & \sim & G_{L}\left[T/E_{C}\right]^{\frac{2}{3}}~.\label{electric_cond}
\end{eqnarray}

{\color{black} We compute the perturbative contribution to the thermoelectric coefficient $G_T$ proportional to $|r|^{2}$ \cite{suppl}
with $\log$-accuracy using three 
parametrizations of the charge, pseudo-spin and flavour modes and symmetrize over three QPC index permutations (re-numerations) \cite{GMb}.
Finally, each QPC contributes equally to $G_{T}$:
\begin{eqnarray}
G_{T}\sim\frac{G_L}{e}|r|^{2}\!\sin\left(2\pi N\right)\!\left[1\!+\!a\cos\left(2\pi N\right)\right]\!\left[\!\frac{T}{E_C}\!\right]\!\!\ln\!\left[\!\frac{E_{C}}{T}\!\right].\!\!\!\;\;\;\;\;\label{lead}
\end{eqnarray}
with $a$$\sim $$1$ \cite{suppl}. Substituting Eqs. (\ref{lead}) and asymptotic equation for $G$ Eq.(\ref{electric_cond}) into the definition of the TP $S=G_{T}/G$ we obtain \cite{footnote4}: 
\begin{eqnarray}
S\sim\frac{1}{e}|r|^{2}\!\sin\left(2\pi N\right)\!\left[1\!+\!a\cos\left(2\pi N\right)\right]\!\left[\!\frac{T}{E_C}\!\right]^{\frac{1}{3}}\!\!\ln\!\left[\!\frac{E_{C}}{T}\!\right].\!\!\!\;\;\;\;\;\label{eq:final_S}
\end{eqnarray}
} 
The perturbative 3CK results for $G_T$ (\ref{lead}) and TP (\ref{eq:final_S}) do not diverge at the limit 
$T$$ \to $$0$ in contrast to 2CK predictions \cite{MA_theory}. Besides, the temperature scaling
of TP $S^{\rm 3CK}$$\propto $$ T^{1/3}\log T$ is consistent with corresponding {\it non-perturbative} scaling 
of the TP maximums $S^{\rm 2CK}_{\rm max}$$ \propto $$ T^{1/2}\log T$ for 2CK. In both cases $S$ vanishes when $T$$\to $$0$. We therefore expect that the scaling (\ref{eq:final_S}) will survive at the limit $T$$\to $$0$ and acquire only marginal modifications in the argument of $\log$ \cite{MA_theory}. 
Eqs. (\ref{lead}-\ref{eq:final_S}) represent the central result of this {\it Letter}.

\textit{Channel symmetry breaking --} {\color{black} We comment
on possible ways to crossover $3CK\to2CK$ and $3CK\to1CK$ in the
charge Kondo circuits. These crossovers have been experimentally
reported in \cite{2CK_experiment_nature2015,3CK_experiment_science2018} and numerically reproduced in \cite{Sela1,Sela2,Sela3} by
using Numerical Renormalization Group (NRG) technique. The simplest
way to describe continuous crossover $3CK\to2CK$ is to imbalance
e.g. the reflection amplitudes in QPC1 and QPC3 \cite{brk}. Having $a_{13}\equiv||r_{1}|-|r_{3}||$
as a relevant perturbation to the symmetric state characterized by
$s_{13}\equiv(|r_{1}|+|r_{3}|)/2\approx|r|$ provides a condition
for a crossover $a_{13}\sim s_{13}$ similar to theory of channel
symmetry breaking of $2CK\to1CK$ discussed in \cite{nkk_2010}. In addition,
the condition $a_{13}s_{13}\ll|r_{2}|^{2}\approx|r|^{2}$ is required.
However, one needs to go beyond the perturbation theory for the quantitative
description of the crossover. The mechanism of $3CK\to1CK$ is more
delicate. First of all, the experiment \cite{3CK_experiment_science2018} shows the non-monotonous behaviour
of conductance evolution  confirmed by non-monotonous NRG flow
in numerical calculations \cite{Sela1, Sela2, Sela3}. Second, the crossover regime has to be
fine tuned by the condition $|a_{13}s_{13}-|r_{2}|^{2}|\ll|r_{2}|^{2}$.
Discussion of these regimes goes beyond the scope of this paper
and will be published elsewhere \cite{tbp}. }

\textit{Discussion and open questions --} {\color{black} 
Describing the quantum thermoelectricity in the NFL regime of the MCK model at the 
strong coupling limit $T\ll T^{\ast}$ is  one of the main open questions.
In particular, it is important to understand if there exists a re-(para)fermionization procedure for $Z_{3}$ fixed point similar to Emery-Kivelson (EK) approach \cite{Emery_Kivelson} developed for $U(1)$$\to $$Z_2$ symmetry reduction \cite{nk2015}. The EK re-fermionization being a cornerstone for understanding of the emergence of the NFL state of 2CK is known to allow straightforward re-formulation of the strong coupling Hamiltonian in terms of $Z_2$ Majorana (para)fermions. However, even if such a procedure does exist for $Z_{3}$ low temperature fixed point \cite{anyons}, the strong coupling Hamiltonian
will not be quadratic anymore in terms of the $Z_3$ parafermions \cite{gefen_1}. Therefore, the non-perturbative treatment of the 3CK problem at its strong coupling will require some additional assumptions or approximations. Yet another challenging question is related to the generalization of the approach developed in this {\it Letter} for the description of the $M>3$ MCK effect at the strong coupling. We expect that even- and odd-  $M$-channel models behave significantly differently: while the ground state of the even-$M=2k$
channel models can be represented in terms of the  Majorana fermions \cite{shura},  $Z_{2k+1}$ parafermions are needed for the description of the odd - $M=2k+1$-channel Kondo physics. Besides, switching between $Z_{2k+1}$ and $Z_{2k}$ low temperature fixed points opens an interesting possibility for investigation the {\color{black} crossovers} between the states with different parafermion fractionalized zero modes. The same goal can be achieved by using the quantum simulators containing a tunnel contact between two different NFL states \cite{nk2018}.\\
 } 
\textit{Conclusions --} In this  {\it Letter} we  address theoretically a fundamental question of the pre-fractionalized zero modes influence on the quantum thermoelectricity of the nano-devices. Using asymptotically exact analytic approach based on abelian bosonization we predict the fractional $T^{1/3}\log T$ low-temperature scaling behaviour of the Seebeck coefficient. While this scaling is obtained perturbatively at ``weak NFL" regime, we present convincing arguments on the validity of the results also at the strong coupling limit. The fractional scaling of the {\it quantum thermoelectric transport} coefficients is closely related to behaviour of {\it quantum thermodynamic} observables \cite{ent1}. We propose to use experimental technique \cite{3CK_experiment_science2018} providing the circuit implementation of quantum simulators of the MCK model for investigation of the parafermion contribution to the quantum thermoelectricity controlled by switching the quantum regimes between different low temperature fixed points.

\textit{Note added} {\color{black} --Recently, a preprint on thermoelectrics of 2CK \cite{DMF} considering closely related problem was posted in the cond-mat archive.}

\textit{Acknowledgements} -- We are grateful to Leonid Glazman, Yuval
Gefen, Joel Moore, Yuval Oreg and Alexander Nersesyan for illuminating discussions.
{\color{black} We especially grateful to Frederic Pierre for the detailed explanation of the experiments \cite{2CK_experiment_nature2015,3CK_experiment_science2018}, 
suggestions \cite{alter_heat} for the experimental realization of the thermoelectric measurements with quantum simulators \cite{fp_private} and discussion of the dynamical Coulomb blockade effects \cite{footnote2,cond_scal}.}
We acknowledge the warm hospitality of the Center for Theoretical Physics
of Complex Systems (PCS) of the Institute for Basic Science (IBS)
and the International Centre of Physics (ICP) of the Institute of Physics
(IOP - VAST). {\color{black} T.K.T.N. acknowledges support through the ICTP Associate Program, ICTP Asian Network on Condensed Matter and Complex Systems, and International Centre of Physics under grant number ICP.2020.03. This research in Hanoi is funded by Vietnam National Foundation for Science and Technology
Development (NAFOSTED) under grant number 103.01-2020.05.}
The work of MK was performed in part at Aspen Center for Physics,
which is supported by National Science Foundation grant PHY-1607611
and partially supported by a grant from the Simons Foundation.



\begin{thebibliography}{99}
\bibitem{alter_heat} {\color{black} Alternatively, the heating of the left electrode (Fig. \ref{f1}) can be achieved by injection of some energy with the voltage biased QPC located upstream the tunnel contact \cite{en_rel, fp_private}. The electrons relax after some adjustable propagation distance to a ``hot" Fermi distribution characterized by the temperature $T+\Delta T$ measured by the noise thermometry \cite{quant_1}}.
\bibitem{comaxes} {\color{black} The integration in $S_0$ is performed along three independent axes $x_1$, $x_2$, $x_3$ (see Fig. \ref{f1}).
The $x$- axis label is dummy and therefore will be omitted. }
\bibitem{comcb} {\color{black}  The fluctuating charge (in units of e) in the
dot area is equal to 
$\pi^{-1}$$\int_{-\infty}^{0}$$\sum_{\alpha=1}^3$$\partial_x$$\phi_\alpha(x,t)$$
dx$$\to $$\pi^{-1}$$\sum_{\alpha=1}^3$$\phi_{\alpha}$$($$0$$,$$t)$ \cite{aleiner_glazman} (only the boundary term is left).}
\bibitem{com_2} {\color{black} In general, for $M$ species of the electrons, the normal modes are charge, spin and $M$$-$$2$ flavour modes. {\color{black} Thus, $M$$-$$1$ modes describing spin and all flavours} are connected to the Cartan (abelian) subgroup of corresponding SU($M$) group \cite{kisbook}.}
\bibitem{GMb}{\color{black} Two diagonal Gell-Mann matrices of the fundamental
representation of SU(3) group are denoted as 
$\lambda_3$$=$${\rm diag}[1,-1,0]$$\equiv$$-$$\lambda_s$ and 
$\lambda_8$$=$${\rm diag}[1,1,-2]$$/$$\sqrt{3}$$\equiv$$\lambda_f$.
These matrices form $\lambda$-Cartan basis. We add the unit matrix to describe charge mode: 
$\lambda_0$$=$${\rm diag}[1,1,1]$$\equiv $$\sqrt{3/2}\lambda_c$.
The transformation from the basis labelling QPC ($1$$-$$3$):
$\vec\phi_{123}(x,t)$$=$$(\phi_1\;\phi_2\;\phi_3)^T$ to the basis of charge, pseudo-spin and flavour modes $c$$,$$s$$,f$:
$\vec\phi^\lambda_{csf}(x,t)$$=$$(\phi_c\;\phi_s\;\phi_f)^T$ is given by
$\phi^\lambda_\alpha$$=$$\frac{1}{\sqrt{2}}$
${\rm Tr}[\lambda_\alpha\cdot\vec\phi_{123}]$ 
where $\alpha$$=$$c$$,$$s$$,$$f$. We define two additional basis of the Gell-Mann matrices through the linear combination of the $\lambda$-Cartan basis:
$\mu_3$$=$$(\lambda_3+\sqrt{3}\lambda_8)/2$$\equiv$$\mu_s$,
$\mu_8$$=$$(\sqrt{3}\lambda_3-\lambda_8)/2$$\equiv$$\mu_f$,
$\rho_3$$=$$(\lambda_3-\sqrt{3}\lambda_8)/2$$\equiv$$\rho_s$,
$\rho_8$$=$$(\sqrt{3}\lambda_3+\lambda_8)/2$$\equiv$$-$$\rho_f$. Corresponding transformation to the $csf$ basis is given by 
$\phi^\mu_\alpha$$=$$\frac{1}{\sqrt{2}}$
${\rm Tr}[\mu_\alpha\cdot\vec\phi_{123}]$ and $\phi^\rho_\alpha$$=$$\frac{1}{\sqrt{2}}$
${\rm Tr}[\rho_\alpha\cdot\vec\phi_{123}]$ with $\mu_c$$=$$\rho_c$$=$$\lambda_c$.
The Jacobian for all transformations $|J|$$=$$1$.}
\bibitem{footnote1} Importantly, for general MCK, we define the charge
mode $\phi_{c}(x,t)=\nicefrac{(\phi_{1}(x,t)+\ldots+\phi_{M}(x,t))}{\sqrt{M}}$.
Only the charge mode enters the charging energy action. Therefore,
{\color{black} the correlator} $K_{C}(\tau)=\left(\nicefrac{\pi^{2}T}{M \gamma E_{C}|\sin\left(\pi T\tau\right)|}\right)^{\nicefrac{2}{M}}$.
\bibitem{footnote2} 
{\color{black} The temperature dependence of the electric conductance 
$G\propto (T/E_C)^{2/3}$ given by (\ref{electric_cond}) exactly corresponds to
the dynamical Coulomb blockade scaling \cite{fp_private}. This dependence was experimentally observed recently \cite{cond_scal}. The scaling (\ref{electric_cond}) is consistent with the orthogonality catastrophe \cite{catastrophe,catastrophe1,Mahan} explanation given in Ref. \cite{Furusaki_Matveev}.}
\bibitem{footnote4} {\color{black} The temperature scaling of $G_T$ for 
$M\ge2$ channel Kondo model is, in general, described by
$G_T\propto T^{1-2/M+2/M}\log T\sim T\log T$ behaviour and thus, in $\log$ approximation does not depend on $M$ (up to $M$-dependent pre-factor). The 
$\log T$ behaviour of $G_T$ is associated with the fluctuations of
$M-1$ gapless modes (spin and flavours) and is generic for all $M\ge2$ channel charge Kondo models characterized by the NFL behaviour. The gate voltage oscillating dependence is described by $\sin[2\pi N(V_g)]$ factor accounting for the charge quantization. The TP scaling is given by  $S^{MCK}\propto T^{1-2/M}\log T$. The TP of all MCK models vanishes at low temperature being however strongly enhanced by $\log T$ factor compared to $S^{1CK}\propto T$ FL scaling.}
\bibitem{brk} {\color{black} When the symmetry between QPCs is explicitly broken,
we use the Cartan basis
compatible to the model with broken symmetry.}
\end{thebibliography}
\end{document}